\documentclass[a4paper,11pt,aps,tightenlines,nofootinbib]{revtex4}
\usepackage{xspace}
\usepackage[english]{babel}
\usepackage{amsfonts}
\usepackage{amsmath, amsthm, amssymb, mathrsfs}
\usepackage[dvips]{graphicx}
\usepackage{pdfpages}
\usepackage{setspace}
\usepackage{indentfirst}
\usepackage[margin=2.5cm]{geometry}
\usepackage{bbm}

\usepackage{pstricks}
\usepackage{enumerate}
\usepackage{pstricks-add}
\usepackage{epsfig}
\usepackage{pst-grad}
\usepackage{pst-plot}

\newcommand{\abs}[1]{\left|#1\right|}
\newcommand{\bra}[1]{\left< #1 \right|}

\newcommand{\ket}[1]{\left|#1\right>}

\newcommand{\be}{\begin{equation}}
\newcommand{\ee}{\end{equation}}
\newcommand{\ba}{\begin{array}{c}}
\newcommand{\ea}{\end{array}}

\def\beqr{\begin{eqnarray}}
\def\eeqr{\end{eqnarray}}

\numberwithin{equation}{section}
\numberwithin{thr}{section}
\numberwithin{chr}{section}
\numberwithin{df}{section}

\begin{document}

\title{New scalar constraint operator for loop quantum gravity}

\author{Mehdi Assanioussi}
\email[]{mehdi.assanioussi@fuw.edu.pl}
\author{Jerzy Lewandowski}
\email[]{jerzy.lewandowski@fuw.edu.pl}
\author{Ilkka M\"{a}kinen,}
\email[]{ilkka.makinen@fuw.edu.pl}
\affiliation{Faculty of Physics, University of Warsaw, Pasteura 5, 02-093 Warsaw, Poland\\}

\begin{abstract}
We present a concrete and explicit construction of a new scalar constraint operator for loop quantum gravity. The operator is defined on the recently introduced space of partially diffeomorphism invariant states, and this space is preserved by the action of the operator. To define the Euclidean part of the scalar constraint operator, we propose a specific regularization based on the idea of so-called ''special'' loops. The Lorentzian part of the quantum scalar constraint is merely the curvature operator that has been introduced in an earlier work. Due to the properties of the special loops assignment, the adjoint operator of the non-symmetric constraint operator is densely defined on the partially diffeomorphism invariant Hilbert space. This fact opens up the possibility of defining a symmetric scalar constraint operator as a suitable combination of the original operator and its adjoint. We also show that the algebra of the scalar constraint operators is anomaly free, and describe the structure of the kernel of these operators on a general level.
\end{abstract}

\maketitle

\section{Introduction}

The canonical quantization of general relativity has come a long way since the formulation of the Ashtekar-Barbero variables \cite{A variables, B variables}. As a generally covariant theory, general relativity has its dynamics encoded in constraints. Loop quantum gravity (LQG) \cite{lqgcan1,lqgcan2,lqgcan3,lqgcan4}, that is the incarnation of the mentioned quantization program, succeeded in defining a Hilbert space of kinematical quantum states, and implementing and solving the Gauss constraints, which encode the $SU(2)$ gauge invariance, and the spatial diffeomorphism constraints \cite{Ashtekar:1995zh}. The scalar constraints are technically more involved because of their complicated expression in terms of the Ashtekar-Barbero canonical variables.

The first rigorous proposal of a scalar constraint operator was introduced by T. Thiemann in \cite{Thiemann96a}, based on some concepts discovered by C. Rovelli and L. Smolin in \cite{Rov-Smo}. The construction involves the volume operator \cite{AshtekarLewand98} and uses a mathematical artifact to suppress the non-polynomial character of the constraints in terms of LQG variables. As a result, the constraint operator is gauge invariant and anomaly-free. This operator acts on the Hilbert space of diffeomorphism invariant states, but does not preserve this space due to the presence of the lapse function in the operator.

Recently, a new Hilbert space ${\cal H}^G_{\rm vtx}$ of partially diffeomorphism invariant states was introduced \cite{Lewandowski:2014hza}. In that article, it was shown that upon some changes in the Thiemann's regularization of the scalar constraints, the resulting quantum operator would preserve ${\cal H}^G_{\rm vtx}$. Moreover, the operator would still be anomaly free and there would be possibilities to define a symmetric constraint operator, making discussions of self-adjointness extensions and spectral analysis more accessible.

In the present article, we explicitly implement the scalar constraints for LQG verifying the criteria discussed in \cite{Lewandowski:2014hza}. We base our construction on ideas and concepts introduced in \cite{RovSmo, Thiemann96a, LQG+SF} to deal with the Euclidean part of the constraint, and the use of the curvature operator introduced in \cite{Curvature_op.} to define the Lorentzian part. The article is organized as follows. In section \ref{sec_1} we briefly review the classical Ashtekar formulation of general relativity. In section \ref{sec_2} we review the Hilbert space of LQG, the implementation of $SU(2)$ gauge invariance and the construction of the partially diffeomorphism invariant Hilbert space ${\cal H}^G_{\rm vtx}$. In section \ref{sec_3} we present the regularization of the classical scalar constraint allowing us to define a non-symmetric scalar constraint operator and its adjoint, both densely defined. We discuss the quantum algebra, the possibility of defining a symmetric constraint operator, then the solutions of the quantum scalar constraints; We close in section \ref{sec_4} with some comments and outlooks to future developments.

\section{Classical theory in Ashtekar variables}\label{sec_1}

The $3+1$ Hamiltonian formulation of general relativity, written in terms of the Ashtekar-Barbero variables \cite{A variables, B variables} $(A^i_a, E_i^a)$ (the spatial index $a$ and the $su(2)$ index $i$ take the values $1,2,3$), manifests as a constrained $SU(2)$ gauge theory. The spatial variable $A^i_a$ and its conjugate momentum $E_i^a$, the densitized triad, verify the canonical relations
\begin{align}\label{PB1}
\{A_a^i(x),E^b_j(y)\}\ =&\ k \beta \delta_a^b\delta^i_j\delta(x,y)\\
\{A_a^i(x),A_b^j(y)\}\ =&\ 0\ =\ \{E^a_i(x),E^b_j(y)\}
\end{align}
where $k= 8\pi G$ and $\beta$ is the Immirzi parameter.

The constraints obtained in this formulation consist of the Gauss constraints ${\cal G}_i(x)$ (gauge constraints), spatial diffeomorphism (vector) constraints $C_a(x)$ and scalar constraints $C(x)$. They are first class constraints and can be expressed as follows:
\be \label{Sing.Const.}
\begin{aligned}
{\cal G}_i(x) &=\frac{1}{k\beta} \left(\partial_a E_i^a(x) + \epsilon_{ij}{}^k A^j_a(x) E^a_k(x)\right), \\
C_a(x) &= \frac{1}{k\beta} F_{ab}^i(x) E_i^b(x) - A_a^i(x){\cal G}_i(x) ,\\
C(x) &= \frac{1}{2k\beta^2} \biggl( \frac{\epsilon_{ijk}E^a_i(x)E^b_j(x)F_{ab}^k(x)}{\sqrt{|\det E(x)|}} + \left(1-s \beta^2\right) \sqrt{|\det E(x)|} \,R(x)\biggr),
\end{aligned}
\ee
where $s =1$ in the case of spacetime with Euclidean signature and $s=-1$ in the case of Lorentzian signature, $F_{ab}^i$ the curvature of the connection $A_a^i$, and $R$ is the Ricci scalar of the metric tensor $q_{ab}$ on the $3$-dimensional manifold $\Sigma$ (the relation between $q_{ab}$ and the variable $E^a_i$ is given by $q^{ab} = E^a_iE^b_i/|\det E|$). This form of the scalar constraints was proposed by Domaga\l{}a \cite{DomLew}, and was used also in our recent paper \cite{LQG+SF}. It is an alternative to Thiemann's form of the scalar constraint \cite{Thiemann96a} used in \cite{Lewandowski:2014hza}.

Imposing the constraints \eqref{Sing.Const.} is equivalent to imposing their smeared versions
\begin{align}\label{Const.}
{\cal G}(\Lambda) = \int \limits_\Sigma d^3x\,\Lambda^i(x) {\cal G}_i(x)\ ,\quad \vec{C}(\vec{N}) = \int \limits_\Sigma d^3x\, N^a(x) C_a(x)\ ,\quad C(N) = \int \limits_\Sigma d^3x\,N(x) C(x),
\end{align}
where $\Lambda(x)=\tau_i \Lambda^i(x)$ is an arbitrary $su(2)$ valued smearing function, while $N^a(x)$ and $N(x)$ are arbitrary real valued smearing functions called the shift and lapse respectively.\\

The constraints algebra reads
\be\label{Const.Alg}
\begin{aligned}
\{{\cal G}(\Lambda),{\cal G}(\Lambda')\} &= {\cal G}([\Lambda,\Lambda']), \\
\{{\cal G}(\Lambda),\vec{C}(\vec{N})\} &= -{\cal G}({\cal L}_{\vec{N}}\Lambda) , \\
\{{\cal G}(\Lambda),C(N)\} &=0,
\end{aligned} \qquad
\begin{aligned}
 \{\vec{C}(\vec{M}),\vec{C}(\vec{N})\} &= \vec{C}({\cal L}_{\vec{M}} \vec{N}) ,  \\
\{\vec{C}(\vec{M}),C(N)\} &= C({\cal L}_{\vec{M}} N)\ , \\
\{C(M),C(N)\} &= \vec{C}(q^{ab}[NM_{,b} - NM_{,b}])+{\cal G}(S(A,E)),
\end{aligned}
\ee
where $S(A,E)$ is a certain function of the phase space variables, whose explicit expression is not relevant for this work but can be found in \cite{lqgcan1}.

The quantization program of LQG is a canonical quantization following Dirac's procedure. Namely, the phase space variables are quantized and a Hilbert space of functionals of the configuration variable $A$ is constructed, then classical functions on the phase space are promoted to quantum operators and the constraints are imposed on the quantum level as operators equations in order to determine the physical Hilbert space. In the following section, we briefly present the construction of the Hilbert space in loop quantum gravity along with the implementation of the Gauss and spatial diffeomorphism constraints.

\section{Loop quantum gravity: kinematics}\label{sec_2}

Loop quantum gravity is an attempt to built a background independent quantum theory of gravity, therefore there is no reference to any background metric in defining the classical algebra to be quantized. Also, since the Poisson brackets (\ref{PB1}) are singular, we need to introduce smeared variables, holonomies and fluxes (defined below), obtained by integration of $A$ and $E$ respectively over appropriate submanifolds of $\Sigma$. 

  \subsection{Kinematical Hilbert space}
  
  The kinematical space in LQG is defined as the space of \emph{cylindrical} functions of the variable $A$, i.e., complex valued functions depending on the $su(2)$-valued differential $1$-form $A =\ A^i_a\tau_i\otimes dx^a$, where $\tau_i\in su(2)$ is a basis of $su(2)$, through finitely many parallel transports (holonomies)
  \begin{equation}
  h_e[A]\ =\ {\rm P}\exp\left(-\int_e A \right)\ ,
  \end{equation}
  where $e$ is an oriented finite curve (\emph{edge}) in $\Sigma$. 
  Then a kinematical quantum state $\Psi$ has the form
  \begin{equation}\label{cyl'} 
  \Psi[A]\ =\ \psi(h_{e_1}[A],\ldots ,h_{e_n}[A]) 
  \end{equation} 
  with a function $\psi\ :\ {\rm SU}(2)^n \rightarrow \mathbb{C}$. The set $\gamma:=\{e_1,...,e_n\}$ is called the graph of $\Psi$.\\
  
  The space of all cylindrical functions with a graph $\gamma$ is denoted by ${{\rm Cyl}}_\gamma$ and the space of all cylindrical functions by ${\rm Cyl}$. The kinematical Hilbert space of LQG, ${\cal H}_{\rm kin}$, is defined as the completion of ${\rm Cyl}$ with respect to the norm defined by a natural scalar product \cite{AL-measure}
  \begin{equation}
  {\cal H}_{\rm kin} \ :=\ \overline{{\rm Cyl}}
  \end{equation}
  
  While a connection operator ``$\widehat{A}$'' is not defined, every cylindrical function $\Psi$ also defines a multiplication operator
  \begin{equation}\label{hol-op}
  (\widehat{\Psi(A)}\Psi')[A]\ =\ \Psi[A]\Psi'[A].
  \end{equation}
  
  The derivative operator is the quantum flux operator, obtained by quantization of the flux corresponding to $E$,
  \begin{align}\label{flux-clas}
  {P}_{S,\xi}\ :=\ \int_S \frac{1}{2}dx^b\wedge dx^c\epsilon_{abc}\xi^i(x) {E}^a_i(x) ,
  \end{align}
  through an oriented 2-dimensional surface $S\subset \Sigma$. Here $\xi:S\rightarrow {\rm su(2)}$ is a (generalized) smearing function that may involve parallel transports depending on $A$. The flux operator corresponding to the classical variable \eqref{flux-clas} is then
  \begin{align}\label{flux-op}
  \hat{P}_{S,\xi}\  =\ \frac{k}{2}\sum_{x\in S}\xi^i(x)\sum_{e}\kappa_S(e)\hat{J}_{x,e,i},
  \end{align}
  where $e$ runs through the germs\footnote{A germ beginning at a point $x$ is the set of curves overlapping on a connected initial segment containing $x$.} beginning at $x$, and $\kappa_S(e) = -1,0,1$ depending on whether $e$ goes down, along, or, respectively, up the surface $S$. The operator $\hat{J}_{x,e,i}$ is assigned to a pair $(x,e)$. Its action on the function $\Psi\in {\rm Cyl}$ defined in (\ref{cyl'}), with $e_1$ belonging to the germ $e$, is given by
  \be
  \hat{J}_{x,e,i}\Psi\ =\ i\hbar\frac{d}{d\epsilon}\bigg|_{\epsilon=0}\psi(h_{e}e^{\epsilon\tau_i},h_{e_2},...,h_{e_n}),
  \ee

  \subsection{Gauss \& spatial diffeomorphism constraints}
  
  In order to complete the quantization program, it is necessary to implement the constraints (\ref{Const.}) and solve them. The Gauss constraint operator can be easily defined in terms of fluxes, and its kernel is identified with the space of gauge invariant cylindrical functions
  \be f(A)\ =\ f(g^{-1}Ag + g^{-1}dg),\ \ \  {\rm for \ every}\ \ g \in C^1(\Sigma,{\rm SU(2)}) .\ee
  We denote their algebra (a subalgebra of Cyl) by Cyl$^G$, and the corresponding Hilbert space ${\cal H}^G_{\rm kin}\subset {\cal H}_{\rm kin}$. A dense subspace of ${\cal H}^G_{\rm kin}$  is spanned by the spin network functions. A spin network function is defined by a graph $\gamma$ with half integer (non zero) spins assigned to the edges, and $SU(2)$ invariant tensors (intertwiners) assigned to the vertices. Then the space of all gauge invariant states can be written as the orthogonal sum
  \be\label{decomp} 
  {\cal H}_{\rm kin}^G\ =\ \overline{\bigoplus_{\gamma} {\cal H}^G_\gamma}
  \ee
  where $\gamma$ ranges over all the classes of graphs\footnote{Two graphs $\gamma$ and $\gamma'$ belong to the same class if $\gamma'$ can be obtained from $\gamma$ by a sequence of the following moves: splitting of an edge, connecting two edges, changing a orientation of an edge.}, and ${\cal H}^G_\gamma$ is the Hilbert space defined as the completion of the space ${\rm Cyl}^G_\gamma$ spanned by the spin-network functions of graph $\gamma$.\footnote{An important subtlety is, that given a graph $\gamma$, we define spin-network functions by non-trivial representations of SU(2) assigned to the edges of $\gamma$. In general, Cyl$^G_\gamma$ contains also spin-network functions defined by a graph $\gamma"$ obtained from $\gamma$ by removing one of the edges.}\\
  
  Let us now turn to the vector constraint. Due to the absence of a well defined operator corresponding to the spatial diffeomorphism constraint functional, the construction of a space of diffeomorphism invariant states is achieved through a diffeomorphism averaging procedure \cite{AL-measure}. The elements of each of the sub-spaces ${\cal H}^G_{\gamma}$ are averaged with respect to all the smooth diffeomorphisms Diff$^\infty(\Sigma)$ which map $\gamma$ into analytic graphs. Recall that given a diffeomorphism $f:\Sigma\rightarrow\Sigma$, its induced action $U_f$ on a cylindrical function (\ref{cyl'} ) is
  $$ (U_f\Psi)[A]\ =\ \psi(h_{f(e_1)}[A], ..., h_{f(e_n)}[A])\ .$$
  But since Diff$^\infty(\Sigma)$ is a non-compact set and we do not know any probability measure on it, we have to define the averaging in  ${\rm Cyl}^*$, the algebraic  dual to ${\rm Cyl}$. The resulting space is a Hilbert space of diffeomorphism invariant states, denoted ${\cal H}^G_{\rm Diff}$, with a scalar product naturally inherited from the scalar product on ${\cal H}^G_{\rm kin}$. 
  
  However, we know that a quantum operator corresponding to the scalar constraint $C(N)$ in (\ref{Const.}) would not preserve the Hilbert space ${\cal H}^G_{\rm Diff}$ because of the presence of the lapse function $N$. In other words, an operator $\hat{C}(N)$ is not invariant under spatial diffeomorphisms. This fact raises serious difficulties in the treatment of relevant questions such as self-adjointness, spectral resolution and anomaly-freeness of the constraints algebra.

  A solution to this issue was suggested recently in \cite{Lewandowski:2014hza}. It consists of introducing an {\it intermediate} space, the vertex Hilbert space ${\cal H}^G_{\rm vtx}$. The idea is to construct from  elements of the Hilbert space ${\cal H}^G_{\rm kin}$  partial solutions to the vector constraints, by averaging the elements of each of the sub-spaces ${\cal H}^G_{\gamma}$ with respect to  all the smooth diffeomorphisms Diff$^\infty(\Sigma)_{{\rm Vert}(\gamma)}$ which act trivially in the set of vertices ${\rm Vert}(\gamma)$. Denote by TDiff$^\infty(\Sigma$)$_\gamma$ the subset of Diff$^\infty(\Sigma$) which consists of all diffeomorphisms $f$ such that $f(\gamma)=\gamma$ and $U_f$ acts trivially in  ${\cal H}^G_{\gamma}$, and by Diff$^\infty_\gamma(\Sigma)_{{\rm Vert}(\gamma)}$ the set of elements of Diff$^\infty(\Sigma)_{{\rm Vert}(\gamma)}$ which preserve the analyticity of $\gamma$. The set of the transformations  ${\cal H}_\gamma^G\rightarrow {\cal H}_{\rm kin}$ induced by Diff$^\infty_\gamma(\Sigma)_{{\rm Vert}(\gamma)}$ can be identified with
  \begin{equation} 
  {\rm D}_\gamma := {\rm Diff}^\infty_\gamma(\Sigma)_{{\rm Vert}(\gamma)} / {\rm TDiff}(\Sigma)_\gamma. 
  \end{equation}
  The averaging is defined in  ${\rm Cyl}^*$ through a rigging map
  \begin{align}\label{eta}
  \eta\ :\ {\rm Cyl}_{\gamma}^G\ &\longrightarrow\ {\cal S}^G_{[\gamma]}\subset{\rm Cyl}^*\\ \nonumber
  \ket{\Psi_{\gamma}}\ &\longmapsto\ \eta(\Psi_{\gamma})\ =\ \frac{1}{N_\gamma} \sum_{[f]\in D_\gamma}\bra{U_f\Psi_{\gamma}} , 
  \end{align}
   where $N_\gamma$  is the number of elements of D$_\gamma$ which preserve the graph $\gamma$.

  The resulting $\eta(\Psi_{\gamma})$ is a well defined linear functional on ${\rm Cyl}^G$. One then extends $\eta(\Psi_{\gamma})$ by linearity to the algebraic orthogonal sum (\ref{decomp}), obtaining a map
  \begin{equation}
  \eta:{\cal H}_\text{kin}^G \longrightarrow {\rm Cyl}^*.
  \end{equation}
  The vertex Hilbert space ${\cal H}^G_{\rm vtx}$ is then defined as the completion
  \begin{equation}
  {\cal H}^G_{\rm vtx}\ :=\ \overline{\eta({\rm Cyl}\cap {\cal H}^G_{\rm kin})} =\ \overline{\bigoplus_{[\gamma]}{\cal S}^G_{[\gamma]}} 
  \end{equation}
  under the norm induced by the natural scalar product
  \be
	\left(\eta(\Psi)|\eta(\Psi')\right)_{\rm vtx}\ =\ \eta(\Psi)\left(\eta(\Psi')\right) .
	\ee
  
  Each state in ${\cal S}^G_{[\gamma]}$ is invariant under the action of elements in $\text{Diff}^\omega(\Sigma)_{{\rm Vert}(\gamma)}$. In this sense, those states are partial solutions to the quantum vector constraint. They can become full solutions of the quantum vector constraint by a similar averaging with respect to the remaining diffeomorphisms Diff($\Sigma$)/Diff($\Sigma$)$_{{\rm Vert}(\gamma)}$, forming the space ${\cal H}_\text{Diff}^G$.

\section{Loop quantum gravity: dynamics}\label{sec_3}

The quantization of the scalar constraint we propose is carried out by treating separately the two terms of the constraint $C(N)$, expressed in equation \eqref{Sc.Const.} below. The first term of $C(N)$ (see (\ref{Sing.Const.})), is quantized using the loop prescription introduced in \cite{LQG+SF} to regularize the curvature of the Ashtekar connection, and Thiemann's trick \cite{Thiemann96a} to remove the non-polynomial dependence on the canonical variables, caused by the presence of the factor $1/\sqrt{|\det E(x)|}$. This  is a special case of quantization of this term proposed in \cite{Lewandowski:2014hza}. The new element is a specific, explicit  proposal for the regulator. The second term of $C(N)$ was already regularized and promoted to a quantum operator, the curvature operator \cite{Curvature_op.}, and we will go briefly through the details of its construction below. Using that operator in our definition of the quantum scalar constraint is a true departure from the paper \cite{Lewandowski:2014hza}.

  \subsection{Regularization of the scalar constraint}\label{Regul.}
  
  The starting point is the expression
  \begin{align}\label{Sc.Const.}
  C(N) = \frac{1}{2k\beta^2} \int \limits_\Sigma d^3x\,N(x) \biggl( \frac{\epsilon_{ijk}E^a_i(x)E^b_j(x)F_{ab}^k(x)}{\sqrt{|\det E(x)|}} + \left(1-s \beta^2\right) \sqrt{|\det E(x)|} \,R(x)\biggr),
  \end{align}
  In the case of $s= 1$ (the space-"time" signature $++++$), the choice $\beta =\pm 1$ kills the second term (and corresponds to the original self-dual Ashtekar variables). For that reason we call the first term  the `Euclidean' part, and we call the second term the `Lorenzian' part.
 
  \subsubsection{Euclidean part}
  
  We first consider the Euclidean part. To express it in a non-singular form, we use Thiemann's trick, which consists of using the identity
  \begin{align}\label{Th.Trick.}
  \frac{\epsilon_{ijk}E^a_i(x)E^b_j(x)}{\sqrt{|\det E(x)|}}= \frac{2}{k}\epsilon^{abc} \{A_c^{k}(x), V\},
  \end{align}  
  where $V$ is the volume of $\Sigma$,
  \begin{align}
  V:=\int \limits_\Sigma d^3x \sqrt{|\det E(x)|}.
  \end{align}
  The Euclidean part $C^E(N)$ then takes the form
  \begin{align}\label{Eucl.}
  C^E(N) := \frac{1}{k^2\beta^2} \int \limits_\Sigma d^3x\,N(x) \epsilon^{abc} F_{ab}^k(x) \{A_c^{k}(x), V\},
  \end{align}
  
  The expression (\ref{Eucl.}) is regularized via approximation of the integral by a Riemannian sum over a partition $\mathscr{C}^\epsilon$, with $\epsilon$ being a parameter characterizing the size of the cells $\Delta$ in $\mathscr{C}^\epsilon$, by replacing $N(x)$ with values of $N$ at a point $x_\Delta$ chosen in each cell $\Delta$, and replacing  the connection coefficients with parallel transports along open curves $s_I(\Delta)$ and the curvature coefficients by the holonomies along loops $\alpha_{IJ}(\Delta)$
  \begin{align}\label{Reg.Eucl.}
  C_{\mathscr{C}^\epsilon}^E(N) = -\frac{1}{k^2\beta^2 W_l^2} \sum_{\Delta \in \mathscr{C}^\epsilon} N(x_\Delta) \epsilon^{IJK} \text{Tr} \Bigl(h_{\alpha_{IJ}(\Delta)}^{(l)} h_{s_K(\Delta)}^{(l)} \{h_{s_K(\Delta)}^{(l)\ -1}, V(\Delta)\}\Bigr),
  \end{align}
  where $h^{(l)}$ is the holonomy in a chosen $SU(2)$ representation $l$ and $W_l=i\sqrt{l(l+1)(2l+1)}$ is a normalization factor\footnote{The representation $l$ is left arbitrary in our construction.  In representation $l$, we choose a basis $\tau_i^{(l)} (i=1,2,3)$ of $su(2)$, satisfying \[ {\rm Tr}\,\bigl(\tau_i^{(l)}\bigr) = 0, \qquad {\rm Tr}\,\bigl(\tau_i^{(l)}\tau_k^{(l)}\bigr) = \frac{W_l^2}{3}\delta_{ik}. \]}, the curves $s_I(\Delta)$ and loops $\alpha_{IK}(\Delta)$ are assigned to each cell $\Delta$ such that this functional converges to $C^E(N)$ in the limit $\epsilon \rightarrow 0$. Below we propose a specific assignment. But before introducing it in detail, we will remind another important element of the procedure. The first, intermediate, step of the quantization is to define in ${\cal H}^G_{\rm kin}$ a partition dependent  quantum operator $\hat{C}_{\mathscr{C}^\epsilon}^E(N)$. This operator will not have a limit when $\epsilon\rightarrow 0$. Still, by duality we want to obtain a well defined operator on ${\cal H}^G_{\rm vtx}$ that carries the diffeomorphism covariance property of the classical constraint. To accomplish that, we need to adapt our regulator to each graph $\gamma$ and the corresponding subspace ${\cal H}^G_{\gamma}$ independently. We propose in this paper the following prescription:
  \begin{itemize}
   \item $\mathscr{C}^\epsilon$ is a triangulation, i.e. each cell $\Delta$ is a tetrahedron;
   \item each tetrahedron $\Delta$ has at most one node of the graph $\gamma$ as one of its vertices;
   \item each node $v$ of the graph $\gamma$ coincides with a vertex of a tetrahedron $\Delta_v$ and $x_{\Delta_v}=v$;
   \item if $v$ is a node of $\gamma$, then
   \begin{itemize}
    \item $v$ is a vertex of $n_v$ tetrahedra $\Delta_v^i$ saturating the neighborhood of $v$ (i.e the tetrahedra meet at $v$ and compose a closed neighborhood centered at $v$);
    \item the edges of the tetrahedra $\Delta_v^i$ saturating the neighborhood of $v$ do not overlap with the edges of $\gamma$ meeting at $v$, except for one tetrahedron, which we call $\Delta_v^{IJK}$. The tetrahedron $\Delta_v^{IJK}$ is adapted to one chosen ordered triple of edges $(e_I,\ e_J,\ e_K)$ meeting at $v$, i.e., the edges $(s_I,\ s_J,\ s_K)$ of $\Delta_v^{IJK}$ meeting at $v$ are segments of the edges $(e_I,\ e_J,\ e_K)$ of the graph $\gamma$ but do not coincide with them;
    \item to the ordered triple of edges $(s_I,\ s_J,\ s_K)$ meeting at $v$ there are assigned three loops $(\alpha^{IJ},\alpha^{JK},\alpha^{KI})$ oriented according to the order of the triple $(s_I,\ s_J,\ s_K)$;
    \item A loop $\alpha^{IJ}$ verifies the following conditions:
    \begin{enumerate}
     \item[i.] $\alpha^{IJ}$ is an analytic curve;
     \item[ii.] $\alpha^{IJ}$ lies in a surface defined through a canonical choice of coordinates adapted to the edges $(s_I,\ s_J,\ s_K)$ and does not intersect the graph\footnote{We do not show the construction of those coordinates nor the rooting procedure for the loop in this article, but we direct the reader to \cite{Thiemann96a} or \cite{lqgcan2} for the details.} $\gamma$ at any point except at $v$;
     \item[iii.] $\alpha^{IJ}$ is tangent to the two edges $e_I$ and $e_J$ of the graph $\gamma$ at the vertex $v$ up to orders $k_I + 1$ and $k_J + 1$ respectively, where $k_I (\geq 0)$ and $k_J (\geq 0)$ are respectively the orders of tangentiality of $e_I$ and $e_J$ at the node\footnote{The order of tangentiality of an edge $e_I$ incident at a node $v$ is the highest order of tangentiality of the edge $e_I$ with the remaining edges incident at $v$ (see \cite{LQG+SF}).};
     \item[iv.] Denote by $s_{IJ}$ the edge of $\Delta_v^{IJK}$ that links the edges $(s_I,\ s_J)$ to form a triangle of the the tetrahedron $\Delta_v^{IJK}$. The shape of the loop $\alpha^{IJ}$ marries the shape of the triangle $(s_I,\ s_J,\ s_{IJ})$ as good as possible;
    \end{enumerate}
   \end{itemize}
  \end{itemize}
	
  This prescription for the adapted partition is twofold: The first part, which contains all the requirements except the conditions on the loops, coincides with some of the requirements on the partition in Thiemann's approach to regularize the scalar constraint \cite{Thiemann96a}. In addition, in \cite{Thiemann96a} the number $n_v$ is set to be equal to $8$ for any node $v$ of the graph thanks to a specific procedure to construct the saturating structure around $v$. We could adopt the same procedure to fix $n_v$ but it is a priori possible to keep it as a free parameter that is the same for all vertices, hence we drop the $v$ label in rest of the article.
  
  The second part of the above prescription is about the conditions on the loop structure. We use a prescription, first introduced in \cite{LQG+SF}, different from the one in Thiemann's construction in which the loop $\alpha^{IJ}$ coincides with the triangle $(s_I,\ s_J,\ s_{IJ})$ of $\Delta_v^{IJK}$. The whole prescription is diffeomorphism invariant and it makes a loop assigned to a pair of edges unique up to diffeomorphisms. As we will see later, the conditions on the loops also allow to introduce a {\it densely} defined adjoint operator of the non-symmetric scalar constraint operator\footnote{In case of Thiemann's construction, the adjoint operator of the non-symmetric scalar constraint operator is not densely defined.}, thereby providing a way to define a symmetric constraint operator (the key condition is that as in \cite{lqgcan1,Lewandowski:2014hza} the loops do not overlap the given graph). In the rest of the article we refer to those loops as {\it special} loops.
  
  Having the adapted partition, we straightforwardly quantize the expression in (\ref{Reg.Eucl.}) by replacing the Poisson bracket of $h_{s_K(\Delta)}^{-1}$ and $V$ with $1/i\hbar$ times the commutator of the corresponding operators, taking for $\hat V$ the internally regularized volume operator of \cite{AshtekarLewand98},
  \begin{align}
    \hat{V}:= l_p^3 \sum \limits_{x \in \Sigma} \hat{V}_x = l_p^3 \kappa_0 \sum \limits_{x \in \Sigma} \sqrt{\abs{\frac{1}{8 \cdot 3!} \sum \limits_{I,J,K} \epsilon(\dot{e}_I,\dot{e}_J,\dot{e}_K) \epsilon_{ijk} \hat{J}_{x,e_I,i}\hat{J}_{x,e_J,j}\hat{J}_{x,e_K,k}}},
  \end{align}
  where $l_p$ is the Planck length, $\kappa_0$ an overall averaging constant, $e_I$ runs through the set of germs starting at the point $x$, and $\epsilon(\dot{e}_I,\dot{e}_J,\dot{e}_K)= {\rm sgn}[\text{det}(\dot{e}_I,\dot{e}_J,\dot{e}_K)]$. Considering a gauge invariant state $\Psi_\gamma$ with a graph $\gamma$, the resulting operator acts as
  \begin{align}\label{Reg.Eucl.Op.}
    \hat{C}_{\mathscr C^\epsilon}^E(N) \Psi_\gamma &:= \sum \limits_{\Delta \in \mathscr C^\epsilon} \hat{C}_{\Delta}^E(N) \notag \\
    &= -\frac{1}{i\hbar k^2\beta^2 W_l^2} \sum \limits_{\Delta \in \mathscr C^\epsilon} \sum \limits_{v \in \Delta \cap \gamma} N(v) \epsilon^{IJK} \text{Tr}\Bigl(h_{\alpha_{IJ}(\Delta)}^{(l)}  h_{s_K(\Delta)}^{(l)}  [h_{s_K(\Delta)}^{(l)\ -1} , \hat{V}]\Bigr) \Psi_\gamma.
  \end{align}
  
  At this stage, the operator defined in (\ref{Reg.Eucl.Op.}) still depends on the triangulation $\mathscr C^\epsilon$. The dependence on the triangulation is removed in three steps:
  \begin{itemize}
    \item[a)] Denote by $R(v)$ the closed region formed by the $n$ tetrahedra $\Delta_v^{(...)}$ of $\mathscr C^\epsilon$ saturating a vertex $v$. Here $(...)$ contains the labels of the edges intersecting at $v$ and defining a specific tetrahedron. Classically, as we take the limit $\epsilon \rightarrow 0$ in the sense of refining the adapted triangulation $\mathscr C^\epsilon$ to another adapted triangulation $\mathscr C^{\epsilon'}$ such that $\epsilon'<\epsilon$, we have
  \begin{align}
    \int \limits_{R(v)} \ \approx \ n \int \limits_{\Delta_v^{IJK}},
  \end{align}
  the label $IJK$ refers to one tetrahedron of $R(v)$. In other words, the integral over $R(v)$ converges to $n$ times the integral over any tetrahedron of $R(v)$ as we take the limit $\epsilon\rightarrow 0$. For the operator in (\ref{Reg.Eucl.Op.}), this translates as
  \begin{align}\label{Reg.Eucl.Op.1}
    \hat{C}_{\mathscr C^\epsilon}^E(N) \Psi_\gamma := -\frac{n}{i\hbar k^2\beta^2 W_l^2} \sum \limits_{v \in \gamma\cap \mathscr C^\epsilon} \sum \limits_{\Delta_v^{IJK} \in \mathscr C^\epsilon} N(v) \epsilon^{IJK} \text{Tr}\Bigl(h_{\alpha_{IJ}(\Delta_v^{IJK})}^{(l)}  h_{s_K(\Delta_v^{IJK})}^{(l)}  [h_{s_K(\Delta_v^{IJK})}^{(l)\ -1} , \hat{V}]\Bigr) \Psi_\gamma .
  \end{align}  
  
  \item[b)] A triangulation $\mathscr C^\epsilon$ selects at each node $v$ of a graph $\gamma$ a unique triple of edges $(e_I,\ e_J,\ e_K)$ meeting at $v$. In order to remove this selection from the operator, it is enough to average at each node $v$ over the classes of triangulations that select different triples meeting at $v$. Therefore the operator would contain contributions from all possible triples meeting at the same node and we obtain
  \begin{align}\label{Reg.Eucl.Op.2}
    \hat{C}_{\epsilon}^E(N) \Psi_\gamma :&= -\frac{n}{i\hbar k^2\beta^2 W_l^2} \sum \limits_{v \in \gamma} \frac{N(v)}{E(v)} \epsilon^{IJK} \text{Tr}\Bigl(h_{\alpha_{IJ}(\Delta)}^{(l)}  h_{s_K(\Delta)}^{(l)}  [h_{s_K(\Delta)}^{(l)\ -1} , \hat{V}]\Bigr) \Psi_\gamma \\ 
    \nonumber &=: \sum \limits_{v \in \gamma} N(v) \hat{C}_{\epsilon, v}^E \Psi_\gamma,
  \end{align}
  where now the $IJK$ run through all triples of edges of the graph $\gamma$ meeting at the node $v$, and $E(v)$ is the number of unordered triples of edges meeting at $v$ (hence $E(v)$ depends only on the graph $\gamma$). Notice that due to the presence of the volume operator in its expression, $\hat{C}_{\epsilon,v}^E$ annihilates two-valent nodes and nodes which have degenerate differential graph structure. Therefore the action of the operator $\hat{C}_{\epsilon}^E(N)$ on a gauge invariant state is always finite and it also preserves the gauge invariant space.
  
  \item[c)] The only dependence left on the triangulation is in $\epsilon$. We then need to take the limit $\epsilon \rightarrow 0$. As we have mentioned above, in this limit $\hat{C}_{\epsilon}^E(N)$ does not converge to any well defined operator in the space ${\cal H}_\text{kin}^G$. The way around this problem is to first pass the operator $\hat{C}_{\epsilon}^E(N)$ to the space ${\cal H}_\text{vtx}^G$ by duality, then take the limit \cite{Lewandowski:2014hza}. The convergence is ensured and the final operator is then defined as
  \begin{align}\label{Eucl.Op.}
    \hat{C}^E(N) := \lim_{\epsilon \rightarrow 0} \left[\hat{C}_{\epsilon}^E(N)\right]^* ,
  \end{align}  
  acting in the space of gauge and partially diffeomorphism invariant states ${\cal H}_\text{vtx}^G$.
  \end{itemize}
  The operator $\hat{C}^E(N)$ is densely defined on the space ${\cal H}_\text{vtx}^G$, as it contains the span of partially diffeomorphism invariant spin networks space $\eta(\mathscr{S})$, and graph changing as it removes special loops at the nodes\footnote{The operator $\hat{C}_{\epsilon}^E(N)$ is regularized in the space ${\cal H}_\text{kin}^G$ and it changes the graph of a state by adding special loops at the nodes. Therefore, the dual operator $\hat{C}^E(N)$ acting ${\cal H}_\text{vtx}^G$ is removing special loops at the nodes.}. It maps its domain ${\mathscr D^E} \subset {\cal H}_\text{vtx}^G$ to a subset of ${\cal H}_\text{vtx}^G$ and therefore preserves the gauge and partial diffeomorphism invariance.\\
  
  \subsubsection{Lorentzian part}
  
  Now let us turn to the Lorentzian part of (\ref{Sc.Const.}), namely
  \begin{align}\label{Loren.}
    C^L(N) = \frac{1- s\beta^2}{2k\beta^2} \int \limits_\Sigma d^3x\,N(x) \sqrt{|\det E(x)|} \,R(x).
  \end{align}
  The quantization of this classical functional was already carried out in \cite{Curvature_op.}. The regularization is external and based on the Regge approximation \cite{Regge1} of the $3$d Einstein-Hilbert action. On a gauge invariant state $\Psi_\gamma$, the non-symmetric operator\footnote{It was shown in \cite{Curvature_op.} that it is possible to obtain a self-adjoint curvature operator from the non-symmetric operator.} corresponding to the Lorentzian part acts as
  \begin{align}
    \hat{C}^L(N) \Psi_\gamma &= \frac{1- s\beta^2}{8k\beta^2} \sum \limits_{v \in \gamma} N(v) \kappa(v)  \sum \limits_{I\neq J} \sqrt{\widehat{V^{-1}}}\hat Y_{e_I,e_J} \sqrt{\widehat{V^{-1}}} \hat\Theta_{e_I,e_J} \notag \\
    &=: \sum \limits_{v \in \gamma} N(v) \hat{C}_v^L \Psi_\gamma,
  \end{align}
  where
  \begin{align}
    \hat Y_{e_I,e_J} &= \sqrt{ (\epsilon_{ijk} \hat{J}_{v,e_I,j} \hat{J}_{v,e_J,k}) (\epsilon_{i j' k'} \hat{J}_{v,e_I,j'} \hat{J}_{v,e_J,k'})}, \\
    \hat\Theta_{e_I,e_J} &= \frac{2\pi}{\lambda_{IJ}} - \pi + \arccos \left[ \frac{ \hat{J}_{v,e_I,j} \hat{J}_{v,e_J,j}}{\sqrt{\hat{J}_{v,e_I,k} \hat{J}_{v,e_I,k}} \sqrt{ \hat{J}_{v,e_J,l} \hat{J}_{v,e_J,l}}} \right],
  \end{align}
  where $\kappa(v)$ is an averaging coefficient that depends only on the valence of the node $v$, $\lambda_{IJ}$ is a free integer parameter \cite{Curvature_op.}, and $\widehat{V^{-1}}$ is the ``inverse volume'' operator defined as
  \begin{align}
    \widehat{V^{-1}}:= \lim \limits_{t \rightarrow 0} \bigl(\hat{V}^2 + t^2 l_p^6 \bigr)^{-1} \hat{V}.
  \end{align}
  The operator $\hat{C}^L(N)$ is not graph changing and passes naturally to the space ${\cal H}_\text{vtx}^G$. It maps its dense domain ${\mathscr D^L} \subset {\cal H}_\text{vtx}^G$ to a subset of ${\cal H}_\text{vtx}^G$ and therefore preserves the gauge and partial diffeomorphism invariance.

  \subsection{Quantum constraints algebra, symmetric constraint operator \& physical states}
  
  We can now introduce the non-symmetric scalar constraint operator
  \begin{align}\label{Const.Op.}
    \hat{C}(N):= \hat{C}^E(N)+\hat{C}^L(N)= \sum \limits_{v \in \gamma} N(v) (\hat{C}_v^E+\hat{C}_v^L) =: \sum \limits_{v \in \gamma} N(v) \hat{C}_v\ .
  \end{align}
  It is defined on a dense domain $\mathscr{D}(\hat{C}(N))\subset{\cal H}_\text{vtx}^G$ and preserves ${\cal H}_\text{vtx}^G$. Since the classical scalar constraint functional is an observable, it is generally assumed that the quantum operator corresponding to it must be self-adjoint. However the operator in (\ref{Const.Op.}) is not symmetric and it was argued in \cite{Non.Sym.Arg.} that it is not necessary to have a self-adjoint constraint operator exactly because it is a constraint\footnote{It was also shown in \cite{Non.Sym.Arg.} that a symmetric constraint operator may not be anomaly free.}. Since we will be looking for the kernel of the scalar constraint operator, it may not be relevant to construct a self-adjoint operator as long as zero belongs to its spectrum. We will show below how we could introduce a symmetric constraint operator which is the first step toward defining a self-adjoint operator. 
  
  \subsubsection{Quantum constraints algebra}
  
   Let us for the moment assume that our constraint operator is $\hat{C}(N)$, then we can make a short calculation to check if this operator is anomaly free. The calculation goes as follows: given a state $\Psi_\gamma \in {\cal H}_\text{vtx}^G$, we have
   \begin{align}
     [\hat{C}(N),\hat{C}(M)] \Psi_\gamma= \sum \limits_{v,v' \in \gamma} N(v)M(v') [\hat{C}_v,\hat{C}_{v'}]\Psi_\gamma\ .
   \end{align}
   Because the regularization used to construct the operator is local with respect to each node, the commutator
   \begin{align}
     [\hat{C}_v,\hat{C}_{v'}] \Psi_\gamma =0 \ ,\ \forall\ v\neq v',
   \end{align}
   hence
   \begin{align}
    [\hat{C}(N),\hat{C}(M)] \Psi_\gamma = \sum \limits_{v \in \gamma} N(v)M(v) [\hat{C}_v,\hat{C}_{v}] \Psi_\gamma\ .
  \end{align}
  
   In the space ${\cal H}_\text{vtx}^G$ the commutator $[\hat{C}_v,\hat{C}_{v}]$ also vanishes for the same reason in the case of Thiemann's constraint operator, namely the two terms of the last commutator, when acting on a state in ${\cal H}_\text{kin}^G$ (before taking the limits of the regulators \ref{Eucl.Op.}), produce two diffeomorphism equivalent states, therefore the commutator vanishes\footnote{The commutator vanishes with respect to URST (topology) \cite{lqgcan2, Rov-Smo}.} on any state in ${\cal H}_\text{vtx}^G$
   \begin{align}
     [\hat{C}(N),\hat{C}(M)]= 0\ .
   \end{align}
  
   When it comes to the algebra with respect to the other constraints, we already know that, on one hand, the operator $\hat{C}(N)$ preserves the $SU(2)$ gauge invariance, on the other hand, a diffeomorphism constraint operator does not exist in this representation and the only thing we could check is whether it is covariant with respect to the action of  diffeomorphisms. The calculation and the result is not different than in the case of Thiemann's constraint operator and we find that indeed the operator $\hat{C}(N)$ is diffeomorphism covariant
   \begin{align}
     U_f^{-1} [\hat{C}(N)] U_f = \hat{C}(f^* N)\ ,\ \forall\ f\in \text{Diff}^\infty(\Sigma)\ .
   \end{align}

   Therefore we conclude that the scalar constraint operator $\hat{C}(N)$ is anomaly free.
  
  \subsubsection{Symmetric scalar constraint operator \& physical states}\label{C_properties}
  
   Concerning the question of defining a symmetric scalar constraint operator, it turns out that it is actually possible to introduce a symmetric operator using $\hat{C}(N)$ and its adjoint operator\footnote{Definition: Let $\hat T$ be a densely defined linear operator on a Hilbert space $\mathscr{H}$. Let $\mathscr{D}(\hat T^\dagger)$ be the set of $\varphi \in \mathscr{H}$ for which there is an $\eta \in \mathscr{H}$ with
   \begin{align}
     \nonumber (\hat T \psi, \varphi) =  (\psi, \eta) \qquad \text{for all } \psi \in \mathscr{D}(\hat T)
   \end{align}
   For each such $\varphi \in \mathscr{D}(\hat T^\dagger)$, we define $\hat T^\dagger \varphi = \eta$. The operator $\hat T^\dagger$ is called the adjoint of $\hat T$.}. The adjoint operator $\hat{C}^\dagger(N)$ is closed and also densely defined ($\mathscr{S}\subset \mathscr{D}(\hat{C}^\dagger(N))$), hence the operator $\hat{C}(N)$ is closable\footnote{We keep the same notation for $\hat{C}(N)$ and its closure.} and $(\hat{C}^\dagger(N))^\dagger = \hat{C}(N)$. Therefore in the rest of the article we consider the closure of $\hat{C}(N)$ and $\hat{C}^\dagger(N)$ as being the non-symmetric scalar constraint operators at our disposal.
 
   The operator $\hat{C}^\dagger(N)$ could be by itself considered as a quantization of the classical scalar constraint functional \ref{Sc.Const.} and it could stand as the quantum scalar constraint operator in the theory on the same footing as the operator $\hat{C}(N)$. If the implementation of the scalar constraint is appropriate, then in the semi-classical limit of the theory the expectation values of the operator and its adjoint should coincide, up to small quantum corrections. Hence, both operators are equally good candidates for the scalar constraint operator in the theory. Notice that $\hat{C}^\dagger(N)$ is also anomaly free, i.e. it preserves $SU(2)$ gauge invariance and we have
   \begin{align}
     [\hat{C}^\dagger(N),\hat{C}^\dagger(M)]= 0\ ,\ U_f^{-1} [\hat{C}^\dagger(N)] U_f = \hat{C}^\dagger(f^* N)\ ,\ \forall\ f\in \text{Diff}^\infty(\Sigma)\ .
   \end{align}
   
   In order to construct a symmetric scalar constraint operator $\hat{C}_\text{sym}(N)$, we suggest to define it as a combination of $\hat{C}(N)$ and $\hat{C}^\dagger(N)$. The simplest example is
   \begin{align}
     \hat{C}_\text{sym}(N):=\frac{1}{2}(\hat{C}(N)+\hat{C}^\dagger(N))\ ,\ \mathscr{D}(\hat{C}_\text{sym}(N))=\mathscr{D}(\hat{C}(N))\cap \mathscr{D}(\hat{C}^\dagger(N))\ .
   \end{align}
   It is obvious that this operator is closable, densely defined and anomaly free. The question of existence of self-adjoint extensions is still open. However it is a strongly eligible candidate for the scalar constraint operator in the theory. The structure of its kernel, equivalently the solutions to this constraint in the space ${\cal H}_\text{vtx}^G$, can to some extent be described quite easily. The properties we know so far of the kernel elements of $\hat{C}(N)$ and $\hat{C}^\dagger(N)$ can be summarized as follows:
   \begin{itemize}
    \item every state that is in the kernel of the volume operator $\hat V$ and has coplanar edges at all the veritices of its graph, is in the kernels of $\hat{C}(N)$ and $\hat{C}^\dagger(N)$;
    \item the set of states of non-zero volume\footnote{By a state of non-zero volume we mean any state which is not in the kernel of the volume operator.} in the kernel of $\hat{C}(N)$ contains an infinite number of states that have the form of finite linear combinations of spin network states\footnote{Simple examples of such states can be straightforwardly derived.};
    \item states of non-zero volume that are in the kernel of $\hat{C}(N)^\dagger$ have the form of infinite linear combinations of spin network states;
    \item states of non-zero volume with graphs that do not contain special loops are neither in the kernel of $\hat{C}(N)$ nor the kernel of $\hat{C}^\dagger(N)$.
   \end{itemize}
   With those properties, we can deduce that the kernel of $\hat{C}_\text{sym}(N)$ has the following structure:
   \begin{itemize}
    \item every state that is in the kernel of the volume operator $\hat V$ and has coplanar edges at all the veritices of its graph, is in the kernel of $\hat{C}_\text{sym}(N)$;
    \item states of non-zero volume that are in the kernel of $\hat{C}_\text{sym}(N)$ have the form of infinite linear combinations of spin network states:
    \item states of non-zero volume with graphs that do not contain special loops are not in the kernel of $\hat{C}_\text{sym}(N)$.
   \end{itemize}
   
   Having a scalar constraint operator, as we mentioned before, the construction of physical states is achieved via averaging of the elements of its kernel, subset of ${\cal H}_\text{vtx}^G$, with respect to the rest of diffeomorphisms in Diff($\Sigma$)/Diff($\Sigma$)$_{{\rm Vert}(\gamma)}$.

\section{Comments \& outlooks}\label{sec_4}

In this article, we presented a concrete implementation of the scalar constraint operator in loop quantum gravity. The construction of the Euclidean part of the constraint operator uses a regularization based on the assignment of ''special'' loops \cite{LQG+SF}, while for the Lorentzian part of the constraint we use the curvature operator of \cite{Curvature_op.}. The resulting non-symmetric operator $\hat C(N)$ is densely defined on the Hilbert space of partially diffeomorphism invariant states ${\cal H}_\text{vtx}^G$, introduced in \cite{Lewandowski:2014hza}. The operator $\hat{C}(N)$ is $SU(2)$ gauge invariant and diffeomorphism covariant, it preserves the space ${\cal H}_\text{vtx}^G$ and its algebra is anomaly-free.

Thanks to the properties of the special loops, the adjoint $\hat{C}^\dagger(N)$ is a densely defined operator on ${\cal H}_\text{vtx}^G$, and has the same properties as $\hat C(N)$. It also allows to construct symmetric constraint operators, $\hat{C}_\text{sym}(N)$, as combinations of the operators $\hat C(N)$ and $\hat{C}^\dagger(N)$. The operators $\hat C(N)$, $\hat C^\dagger(N)$ and $\hat{C}_\text{sym}(N)$ are all equally suitable candidates for the scalar constraint operator in loop quantum gravity. In each case, the general structure of the kernel of the constraint operator is known on a qualitative level, as outlined in section \ref{C_properties}.

The regularization proposed in this article could also be applied in order to define a Master constraint operator, corresponding to the classical Master constraint functional introduced by Thiemann \cite{Thiemann98} as a way of reformulating the singular scalar constraints $C(x)$ of equation \eqref{Sing.Const.}. Carrying out the construction, one would obtain a densely defined operator on ${\cal H}_\text{Diff}^G$ which is symmetric, gauge and diffeomorphism invariant, and anomaly-free. However, further work is needed in order to investigate the structure of the kernel of this operator.

The freedom of choice between different eligible scalar constraint operator should be regarded as a quantization ambiguity that can be fixed only through a semi-classical analysis of the dynamics in the theory. Therefore the next step of our program is the challenging task of constructing, or at least approximating, semi-classical states in the theory. The example of the operator $\hat{C}(N)$ is encouraging in this direction since its kernel is more tractable with respect to the spin network basis.

\begin{center}
 \large{\bf{Acknowledgments}}
\end{center}
This work was supported by the grant of Polish Narodowe Centrum Nauki nr 2011/02/A/ST2/00300. I.M. would like to thank the Jenny and Antti Wihuri Foundation for support.\\

\vspace{0cm}

\bibliographystyle{plainnat}

\end{document}